\documentclass[prl,twocolumn]{revtex4}
\usepackage{graphicx}
\usepackage{amssymb,amsmath}
\usepackage{enumerate}

\newcommand{\be}{\begin{eqnarray}}
\newcommand{\ee}{\end{eqnarray}}

\begin{document}

\title{Distributed chaos in the Hamiltonian dynamical systems and isotropic homogeneous turbulence}

\author{A. Bershadskii}

\affiliation{
ICAR, P.O. Box 31155, Jerusalem 91000, Israel
}

\begin{abstract}
It is shown that the distributed chaos in the simple Hamiltonian (conservative) dynamical systems, such as the Nose-Hoover oscillator and double oscillator, can mimic the distributed chaos in the isotropic homogeneous turbulence. Direct numerical simulations with the classic Toda lattice and with the nonlinear Schr\"{o}dinger equation (soliton turbulence) under random initial conditions have been also discussed in this context. These properties of the distributed chaos have been related to analytical properties of the Hamiltonian systems. Decrease of the smoothness results in the power-law spectra instead of the stretched exponential ones characteristic to the distributed chaos, both for the dynamical systems and for turbulence.

\end{abstract}

\maketitle

\section{Distributed chaos in smooth systems}
\begin{figure}
\begin{center}
\includegraphics[width=8cm \vspace{-0.3cm}]{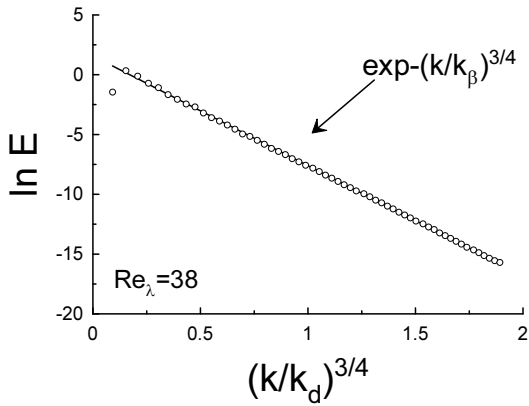}\vspace{-4cm}
\caption{\label{fig1} 3D energy spectrum for isotropic homogeneous (steady) turbulence \cite{gfn}. The straight line indicates the stretched exponential decay Eq. (19).} 
\end{center}
\end{figure}
 It is well known that smooth chaotic dynamical systems often exhibit exponential power spectra (see, for instance \cite{sw}-\cite{sig} and references therein):
$$
 E(f) \propto \exp-(f/f_c) \eqno{(1)}
$$ 
  For Hamiltonian dynamical systems it is not a generic case. For instance, a spontaneous time translational symmetry breaking with action as an adiabatic invariant (instead of energy \cite{suz}, due to the Noether theorem relating the energy conservation to the time translational symmetry) produces a distributed chaos with a stretched exponential spectrum \cite{b1}:
$$
 E(f) \propto \exp-(f/f_0)^{1/2} \eqno{(2)}
$$   
This stretched exponential comes from representation of the power spectrum as a weighted superposition of the exponentials 
$$
E(f ) \propto \int_0^{\infty} P(f_c) \exp -(f/f_c)~ df_c  \eqno{(3)}
$$
(the name: distributed chaos, came because of this representation).  
    To show this let us, following to the Ref. \cite{b1}, consider relationship between characteristic velocity $v_c$ and the characteristic frequency $f_c$ \cite{suz}
$$    
v_c \propto I^{1/2} f_c^{1/2}  \eqno{(4)}
$$
where $I$ is the action. For a Gaussian distribution of the characteristic velocity $v_c$ one obtains the chi-squared ($\chi^{2}$) distribution for $f_c$:
$$
P(f_c) \propto f_c^{-1/2} \exp-(f_c/4f_0)  \eqno{(5)}
$$
where $f_0$ is a constant. Then, substituting Eq. (5) into Eq. (3) one obtains the stretched exponential spectrum Eq. (2).\\

  This consideration can be extended also on the time independent smooth Hamiltonian. Let us consider an analytical expansion at frequency $f=0$ of power spectrum for the position $q(t)$
$$
E_q(f) = I_q + I_q^{(1)}f +I_q^{(2)}f^2 +......   \eqno{(6)}
$$
and analogously for the momentum $p(t)$ 
$$
E_p(f) = I_p + I_p^{(1)}f +I_p^{(2)}f^2 +......   \eqno{(7)}
$$
For the invariant action $I$
$$
I_p = c\cdot I \eqno{(8)}
$$
$c$ is a dimensionless constant. This relationship allows consider  $I_p$ as a momentum-action and $I_q$ as a position-action. \\
  
  From the dimensional considerations one can estimate
$$
v_c \propto I_p^{1/2} f_c^{1/2}  \eqno{(10)}
$$
instead of Eq. (4), or
$$  
v_c   \propto I_q^{1/2} f_c^{3/2}  \eqno{(11)}
$$
In a general form
$$
v_c \propto f_c^{\alpha}  \eqno{(12)}
$$

For a stretched exponential spectrum 
$$
E(f ) \propto \int_0^{\infty} P(f_c) \exp -(f/f_c)df_c  \propto \exp-(f/f_0)^{\beta}  \eqno{(13)}
$$
the $P(f_c)$ asymptote at $f_c \rightarrow \infty$ is 
$$
P(f_c) \propto f_c^{-1 + \beta/[2(1-\beta)]}~\exp(-bf_c^{\beta/(1-\beta)}) \eqno{(14)}
$$
$b$ is a constant \cite{jon}. 

 Probability distributions $\mathcal{P} (v_c)$ and $P(f_c)$ can be related as
$$
\mathcal{P} (v_c) dv_c \propto P(f_c) df_c  \eqno{(15)}
$$
Then from the Eqs (12) and (15):
$$
P(f_c)  \propto f_c^{\alpha -1} ~\mathcal{P} (v_c(f_c)) \eqno{(16)}
$$ 
 
  It follows from the  Eqs. (12),(14) and (16) that for the Gaussian $\mathcal{P} (v_c)$:
$$
\beta = \frac{2\alpha}{1+2\alpha}   \eqno{(17)},
$$
i.e. for $\alpha = 1/2$ (Eq. (10)) the parameter $\beta =1/2$ - Eq. (2), and for $\alpha = 3/2$ (Eq. (11)) parameter $\beta =3/4$:
$$
E(f) \propto \exp-(f/f_0)^{3/4} \eqno{(18)}
$$ 

   For isotropic homogeneous turbulence the Birkhoff-Saffman invariant \cite{bir},\cite{saf} (related to the space translational symmetry - homogeneity) provides the distributed chaos with the stretched exponential spectrum \cite{b2}
$$
 E(k) \propto \exp-(k/k_0)^{3/4} \eqno{(19)}
$$ 
where $k$ is wavenumber. 

  Figure 1 shows 3D energy spectrum obtained in a
direct numerical simulation of isotropic and homogeneous steady 3D turbulence \cite{gfn} for the Taylor-Reynolds number $Re_{\lambda} = 38$, i.e. at the onset of the turbulence (about this value of $Re_{\lambda}$ cf. Eq. (10) of the Ref. \cite{sb} and the Ref. \cite{s}). The straight line indicates (in the scales chosen for this figure) the stretched exponential decay Eq. (19). 
\begin{figure}
\begin{center}
\includegraphics[width=8cm \vspace{-1cm}]{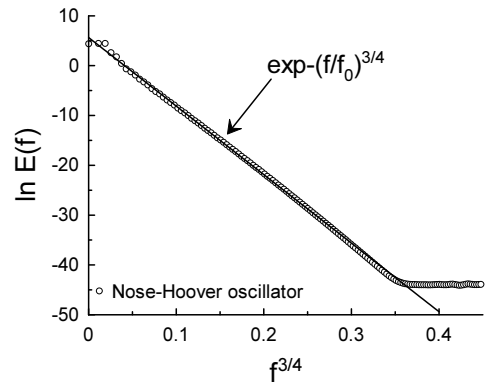}\vspace{-4cm}
\caption{\label{fig2} Power spectrum of the $x$ fluctuations for the Nose-Hoover oscillator.}. 
\end{center}
\end{figure}
  Since the Birkhoff-Saffman invariant overcomes the dissipative (viscous) nature of the Navier-Stokes equation \cite{bir},\cite{saf} it is interesting to compare the distributed chaos corresponding to the isotropic homogeneous turbulence with that corresponding to the smooth Hamiltonian dynamical systems: Eqs. (18) and (19).

\section{The Nose-Hoover oscillator}

  The Nose-Hoover oscillator (the Sprott A system \cite{spot}) can be considered as a harmonic oscillator contacting with a thermal bath. This oscillator is described by the system of equations
$$
\left\{ \begin{array}{l} \dot{x} = y \\[0.1cm] \dot{y} = -x + yz \\[0.1cm] \dot{z} = 1 - y^2 \end{array} \right.  \eqno{(20)}
$$
(the overdot denotes a derivative in time). The nonlinear term - yz, represents the thermostat.
   A remarkable property of this system is its Hamiltonian character, with Jacobian matrix
$$  
  {\cal J} = \left[ \begin{array}{ccc} 0 & 1 & 0 \\[0.1cm] -1 & z & y \\[0.1cm] 0 & -2y & 0 \end{array} \right]  \eqno{(21)}
$$ 
and with a trace 
$$
\mathrm{Tr} ~J = z \eqno{(22)}
$$
It can be shown that
$$
\frac{1}{T} \int_{t=0}^T z \, {\rm d}z = 0  \eqno{(23)}
$$
Conservative systems have the rate of the phase space volume expansion equal to the $\mathrm{Tr} ~J = 0$.

  Equations (20) generate time-reversible Hamiltonian (phase-volume conserving) {\it chaotic} solutions for only a small set of initial conditions, while the trajectories generated by most of the other initial conditions have been attracted by invariant tori.

 Figure 2 shows power spectrum computed for the $x$ fluctuations of the Nose-Hoover oscillator. The time series data were taken from the site Ref. \cite{gen} (see also the Ref. \cite{spot}). The maximum entropy method (providing optimal spectral
resolution for the chaotic spectra \cite{oh}) was used for this computation. The straight line in this figure indicates correspondence to the Eq. (18) (cf. Fig. 1).

\section{Double pendulum}

 The double pendulum consists of two point masses ($m_1$ and $m_2$) at the end of massless straight strings (with fixed length: $l_1$ and $l_2$). The two simple regular pendula are joined together for free oscillation in a plane. The angles between each straight string and the vertical are $\theta_1$ and $\theta_2$. These angles are used as generalized coordinates. The dynamical system is Hamiltonian (conservative) and equations of motion can be written as:
$$
(m_1 + m_2)l_1 \ddot{\theta}_1 + m_2l_2\ddot{\theta}_2\cos(\theta_1 - \theta_2) +
$$
$$
+ m_2l_2\dot{\theta}_2^2\sin(\theta_1 - \theta_2) + (m_1+m_2)g\sin\theta_1 = 0, \eqno{(24)}
$$
$$
m_2l_2\ddot{\theta}_2 + m_2l_1\ddot{\theta}_1\cos(\theta_1 - \theta_2) -
$$
$$
- m_2l_1\dot{\theta}_1^2\sin(\theta_1 - \theta_2) + m_2g\sin\theta_2 = 0   \eqno{(25)}
$$
This system under certain conditions exhibits a chaotic behaviour (see, for instance, Ref. \cite{so} and references therein). 

\begin{figure}
\begin{center}
\includegraphics[width=8cm \vspace{-1cm}]{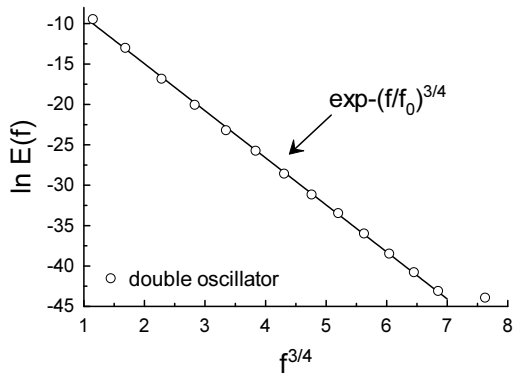}\vspace{-4.2cm}
\caption{\label{fig2} The high-frequency part of the power spectrum of the double oscillator coordinate $\theta_1$ fluctuations. } 
\end{center}
\end{figure}
  
  In the recent Ref. \cite{tar} this system was numerically solved with:
$m_1 = 3,~ m_2 = 1,~ l_1 = 2,~ l_2 = 1,~ g = 1$ and the initial conditions: $  \theta_1(0) = 0,~ \theta_2(0) = \mathrm{arccos}(-7.99/8.0)$. Figure 3 shows a high-frequency part of the broadband power spectrum of the $\theta_1$ fluctuations for this solution in the appropriately chosen scales (cf. Figs. 1,2). The spectral data were taken from the Fig. 2.17A of the Ref. \cite{tar}. The straight line in this figure indicates correspondence to the Eq. (18).

\section{The Toda system}
 
 Chaos and integrability are often considered as mutually excluding and even opposite types of behaviour of the Hamiltonian systems (see, for instance, the Ref. \cite{tab}). The classic many-particle Toda system is a completely integrable and solvable one (see, for instance, Ref. \cite{toda} and references therein). Therefore, results of a numerical experiment with this system (under random initial conditions) reported in the recent Ref. \cite{ebf} are of a special interest.
 
    The Toda system (lattice) is a one dimensional system of $N$ particles with coordinates $q_i$. These particles interact according to the potential $V(q_i,q_{i+1})=c\ e^{-\alpha(q_i-q_{i+1})}$ (with $\alpha$ and $c$ as constants). If $\alpha = c =1$ the dynamic equations describing this system are
$$
\dot{q_i}=p_i,   \eqno{(26)}
$$
$$
\dot{p_i}=e^{q_i-q_{i-1}}-e^{q_{i+1}-q_i}  \eqno{(27)} 
$$
The numerical experiment was performed for $N=32$ particles with random initial conditions and zero total momentum. All the integrals of motion, related to the complete integrability, were found to be approximately constant. However, the power spectrum of the coordinates was found having a broadband high-frequency part, characteristic to the chaotic dynamical systems. Figure 4 shows this high-frequency part for the coordinate $q_1$ in the appropriately chosen scales (cf. Figs. 1-3). The spectral data were taken from the Fig. 10a of the Ref. \cite{ebf}. The straight line indicates correspondence to the Eq. (18).
\begin{figure}
\begin{center}
\includegraphics[width=8cm \vspace{-1.1cm}]{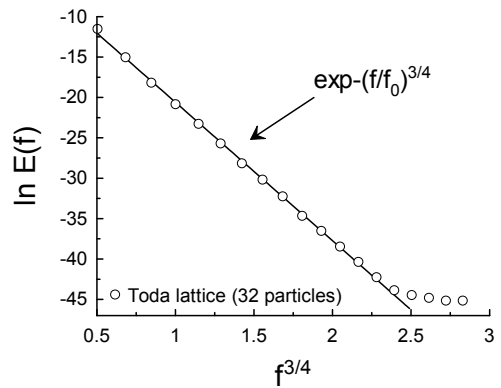}\vspace{-4cm}
\caption{\label{fig2} The high-frequency part of the power spectrum of the Toda lattice coordinate $q_1$ fluctuations for the numerical experiment \cite{ebf}. } 
\end{center}
\end{figure}
 \begin{figure}
\begin{center}
\includegraphics[width=8cm \vspace{-0.9cm}]{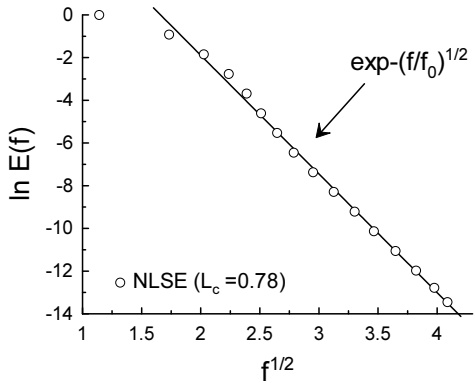}\vspace{-4.3cm}
\caption{\label{fig2} The high-frequency part of the power spectrum for the NLSE direct numerical simulation at the slightly supercritical value of $L_c = 0.78$  and $\xi=100$. } 
\end{center}
\end{figure}  
  
  In order to explain this situation one should recall that for the Toda system any truncation of the exponential potential (using a truncated Taylor series expansion of order higher than two) results in violation of the system integrability \cite{yo}.  Due to this sensitivity, the complete theoretical integrability has a restricted applicability in the numerical experiments: the high-frequency parts of the spectra could be affected by this sensitivity and behave as those of the nonintegrable systems (exhibiting the chaotic features), while their contribution to the integrals will be negligible due to the very fast decay (and the integrals will be still approximately constant).\\

\section{Nonlinear Schr\"{o}dinger equation - soliton dominated chaos}

 Another presumably integrable Hamiltonian dynamical system is described by  the Nonlinear Schr\"{o}dinger equation (NLSE) \cite{zs}. Unlike the previously considered dynamical systems with finite number of the degrees of freedom this system is described by the nonlinear partial differential equation (with infinite number degrees of freedom). The integrability results in the existence of solitons and breathers having considerable amplitude and located in the chaotic sea of the low amplitude radiation waves. The solitons generally move in all directions and collide producing the high amplitude perturbations of the chaotic field. The chaos is deterministic because the NLSE dynamics is entirely determined by the initial conditions. \\

   In a recent Ref. \cite{ahm} results of a direct numerical simulation of the dimensionless NLSE for the envelope of a physical field 
$$
{\rm i} \frac{\partial \psi}{\partial \xi} = - \frac{1}{2}\frac{\partial^2 \psi}{\partial \tau^2} - |\psi|^2 \psi,
\eqno{(28)}
$$   
with random initial conditions on the evolution variable $\xi$ (see Ref. \cite{ahm} for a detail description) and periodic conditions for the transverse variable $\tau$, were reported. 

   In order to remove the breather component of the chaotic field the continuous waves were omitted in the initial conditions. If the correlation length in the initial conditions $L_c$ was small enough, then the chaotic field does not contain solitons as well. The low amplitude radiation waves practically do not change initial physical spectrum (parabolic in the semi-logarithmic scales) during the entire process of evolution.
\begin{figure}
\begin{center}
\includegraphics[width=8cm \vspace{-1.05cm}]{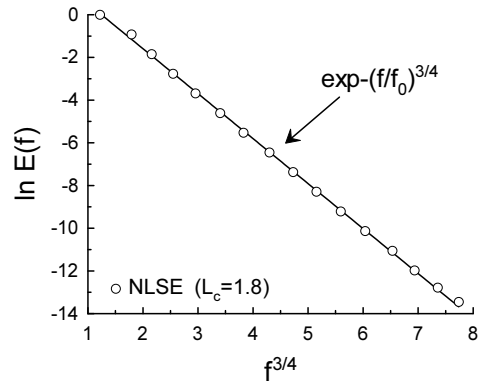}\vspace{-4cm}
\caption{\label{fig2} The high-frequency part of the power spectrum for the NLSE direct numerical simulation at the $L_c = 1.8$ and $\xi=100$ (fully developed soliton chaos). } 
\end{center}
\end{figure}
\begin{figure}
\begin{center}
\includegraphics[width=8cm \vspace{-0.3cm}]{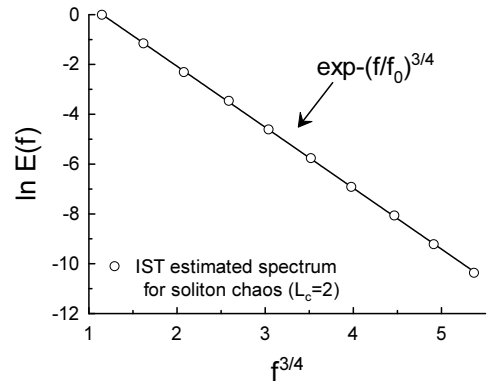}\vspace{-4.3cm}
\caption{\label{fig2} The high-frequency part of the power spectrum estimated by the inverse scattering technique simulation at $L_c = 2$  and $\xi=100$ (fully developed soliton chaos). } 
\end{center}
\end{figure}
The correlation length in the initial conditions had been increased up to a slightly supercritical value $L_c = 0.78$ 
where the  nonlinearity and generation of solitons were activated, resulting in a mixture of the low amplitude radiation waves and the higher amplitude solitons. Figure 5 shows logarithm of the spectrum vs. $f^{1/2}$ (where $f$ denotes the transverse frequency) for this situation at $\xi=100$. According to the authors of the Ref. \cite{ahm} at $\xi=100$ the chaos can be considered as a statistically stationary one. The spectral data for the Fig. 5 were taken from the Fig. 12b of the Ref. \cite{ahm}. The straight line in the Fig. 5 indicates the stretched exponential Eq. (2) for high frequencies.  

 In order to increase number of solitons and, especially, their amplitudes the correlation length in the initial conditions had been further increased to $L_c = 1.8 $. Figure 6 shows the power spectrum for this situation at $\xi=100$. The spectral data were taken from the Fig. 12c of the Ref. \cite{ahm}. The straight line in the figure indicates the stretched exponential Eq. (18) for high frequencies (cf. Fig. 1).  \\

 In the Ref. \cite{ahm} a rough numerical estimation for the spectrum of the soliton dominated chaos was also made using the inverse scattering technique - IST (which is based on the integrability of the NLSE). The authors used a simplified assumption that the spectrum of the soliton dominated chaos can be roughly estimated as a superposition of the spectra of the singular solitons (ignoring contribution of the soliton collisions and of the low amplitude radiation waves). In this approximation the square of the sum of the singular soliton Fourier transforms (obtained by the inverse scattering technique)  corresponds to the overall spectrum of the soliton dominated chaos. The authors of the Ref. \cite{ahm} have the spectra averaged over 26 realizations. Figure 7 shows the spectrum obtained in this simulation at $\xi=100$ and $L_c=2$. The spectral data were taken from the Fig. 11 of the Ref. \cite{ahm}. The straight line in the figure indicates the stretched exponential Eq. (18).\\
 
  It is also interesting to recall the Hasimoto's mapping of the vortex filaments' solutions of the Euler's equation to the soliton solutions of the NLSE \cite{has} (see also the Refs. \cite{sb} and \cite{bkt}).\\
\begin{figure}
\begin{center}
\includegraphics[width=8cm \vspace{-1.1cm}]{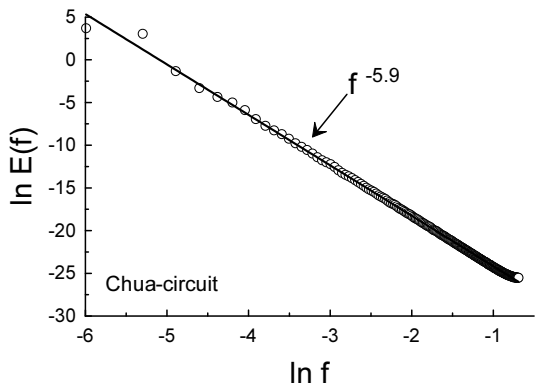}\vspace{-4.18cm}
\caption{ Power spectrum of the $x(t)$ component for the Chua-circuit. } 
\end{center}
\end{figure}
\begin{figure}
\begin{center}
\includegraphics[width=8cm \vspace{-0.7cm}]{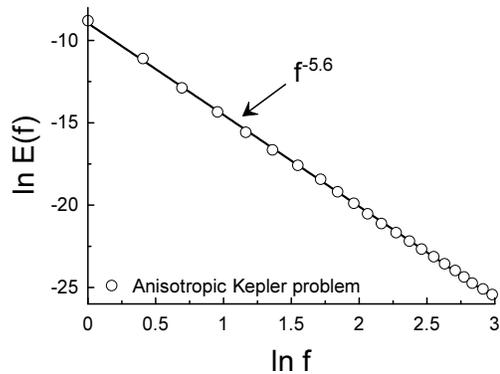}\vspace{-5cm}
\caption{  Power spectrum (a decaying part) of a $z_1$ variable computed for the anisotropic Kepler problem. } 
\end{center}
\end{figure}
\section{Non-smooth systems and power-law spectra}

 As in the first Section let us start from a simple non-Hamiltonian non-smooth dynamical system - Chua-circuit \cite{spot}:
$$
\begin{aligned}
~~~~~~\dot{x}= a(y-h(x)),~~~~~~~~~~~~~~~~~~~~~~~~~~~~~~~~~ \\
~~~~~~~~~~~~\dot{y}= x-y+z,  ~~~~~~~~~~~~~~~~~~~~~~~~~~~~~~(29)  \\
~~~~~~\dot{z}= -by      ~~~~~~~~~~~~~~~~~~~~~~~~~~~~~~~~~~~~~~~~~~~~~
\end{aligned}
$$
where
$$
h(x) = m_1x+\frac{1}{2}(m_0-m_1) (|x+1|-|x-1|)~~~(30)
$$\\

  Figure 8 shows (in the log-log scales) power-law spectral decay
$$
E(f) \propto f^{-\gamma}   \eqno{(31)}
$$
of the $x(t)$ component at $a=9.35$, $b=14.79$, $m_0 = -1/7$, $m_1=2/7$. The time-series data were taken from the site \cite{gen}. Unlike the smooth dynamical systems the non-smooth dynamical system with chaotic dynamics has a power-law spectral decay Eq. (31) indicated in the log-log scales by a straight line corresponding to $\gamma \simeq 5.9$.\\ 
  
  Next example is a non-smooth Hamiltonian system. The anisotropic Kepler problem represents an electron motion in the Coloumb field for an anisotropic crystal \cite{rs}. Electron's effective mass is different along the $z$-direction and in the $x$-$y$ plane. In the anisotropic case the system is non-smooth and non-integrable. 
  
  Figure 9 shows power spectrum computed using the data obtained in the direct numerical simulations of the anisotropic Kepler problem (the spectral data were taken from the Ref. \cite{tar}). The straight line indicates the power-law spectral decay Eq. (31) with $\gamma \simeq 5.6$.\\
  
  Figure 10 shows, in the log-log scales, 3D energy spectrum obtained in the direct numerical simulation of isotropic and homogeneous steady 3D turbulence \cite{gfn} for the Taylor-Reynolds number $Re_{\lambda} = 125$ (cf. Fig. 1). The dashed curve indicates the stretched exponential decay Eq. (19), whereas the straight line with the slope '-5/3' is drawn for reference of the power-law spectral decay $E(k) \propto k^{-5/3}$ (the Kolmogorov spectrum \cite{my}). One can see that with the increase of the $Re_{\lambda}$ the smoothing effect of viscosity becomes insufficient for the small values of the wavenumber and the 'inertial' power-law range appears at these wavenumbers (cf. Ref. \cite{my}).

\section{Acknowledgement}

I thank  T. Nakano, D. Fukayama, T. Gotoh, and  J.C. Sprott for sharing 
their data, and A. Pikovsky for stimulating discussion.

\begin{figure}
\begin{center}
\includegraphics[width=8cm \vspace{-1.2cm}]{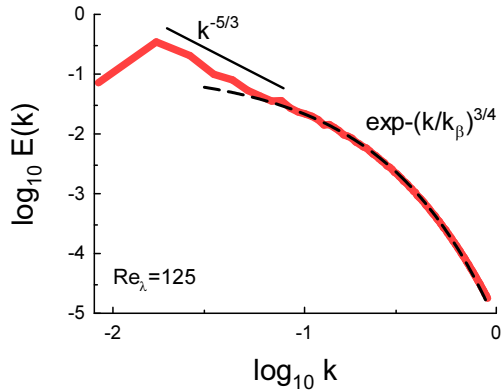}\vspace{-3.7cm}
\caption{ 3D energy spectrum for isotropic homogeneous (steady) turbulence for $Re_{\lambda} = 125$.  } 
\end{center}
\end{figure}

\end{document}